\documentclass[a4paper,fleqn,usenatbib]{mnras}

\usepackage{mathptmx}

\usepackage[T1]{fontenc}
\usepackage{ae,aecompl}

\usepackage{graphicx}	
\usepackage{amsmath}	
\usepackage{amssymb}	

\title[Searching for new white dwarf pulsators for \textit{TESS}]{Searching for new white dwarf pulsators for \textit{TESS} observations at Konkoly Observatory
}

\author[Zs. Bogn\'ar et al.]{
Zs. Bogn\'ar,$^{1}$\thanks{E-mail: bognar.zsofia@csfk.mta.hu (Zs.B.)}
Cs. Kalup,$^{1,2}$
\'A. S\'odor,$^{1}$
S. Charpinet$^{3,4}$
J.~J. Hermes$^{5}$\thanks{Hubble Fellow}
\\
$^{1}$Konkoly Observatory, MTA Research Centre for Astronomy and Earth Sciences, Konkoly Thege Mikl\'os \'ut 15-17, H--1121 Budapest\\
$^{2}$E\"otv\"os University, Department of Astronomy, Pf. 32, 1518, Budapest, Hungary\\
$^{3}$Universit\'e de Toulouse, UPS-OMP, IRAP, Toulouse, F-31400, France\\ 
$^{4}$CNRS, IRAP, 14 avenue Edouard Belin, F-31400 Toulouse, France\\
$^{5}$Department of Physics and Astronomy, University of North Carolina, Chapel Hill, NC -- 27599-3255, USA
}

\date{Accepted XXX. Received YYY; in original form ZZZ}

\pubyear{2018}

\begin{document}
\label{firstpage}
\pagerange{\pageref{firstpage}--\pageref{lastpage}}
\maketitle

\begin{abstract}
We present the results of our survey searching for new white dwarf pulsators for observations by the \textit{TESS} space telescope. We collected photometric time-series data on 14 white dwarf variable-candidates at Konkoly Observatory, and found two new bright ZZ\,Ceti stars, namely EGGR\,120 and WD\,1310+583. We performed the Fourier-analysis of the datasets. In the case of EGGR\,120, which was observed on one night only, we found one significant frequency at 1332\,$\mu$Hz with 2.3\,mmag amplitude. We successfully observed WD\,1310+583 on eight nights, and determined 17 significant frequencies by the whole dataset. Seven of them seem to be independent pulsation modes between 634 and 2740\,$\mu$Hz, and we performed preliminary asteroseismic investigations of the star utilizing six of these periods. We also identified three new light variables on the fields of white dwarf candidates: an eclipsing binary, a candidate delta Scuti/beta Cephei and a candidate W UMa-type star.    

\end{abstract}




\section{Introduction}


The main goal of the \textit{TESS} (Transiting Exoplanet Survey Satellite; \citealt{2015JATIS...1a4003R}) all-sky survey space project, as part of NASA's Explorer programme, is to detect exoplanets at nearby and bright (up to about 15 magnitude) stars, applying the transit method. The telescope is planned to obtain data during a two-year time span, utilizing four CCD cameras, which provide a 24$\times$96 degree field-of-view altogether. It will allow data acquisition with time samplings of 2\,min for selected targets, with the possibility for 20\,s sampling after commissioning. The shorter sampling times make the project feasible to obtain data for the investigations of the light variations of the short-period pulsating white dwarf (WD) and subdwarf stars, too.

In the frame of this ground-based photometric survey presented, we selected for observations white dwarf pulsator-candidates listed by the \textit{TESS} Compact Pulsators Working Group (WG\#8), which do not lie far from the ZZ\,Ceti or V777\,Her instability domains. We note that a more extensive search has been presented by \cite{2017MNRAS.472.4173R}, where they have compiled an all-sky catalogue of ultraviolet, optical and infrared photometry, and presented data for almost 2000 bright white dwarfs and six ZZ Ceti candidates.

ZZ\,Ceti (or DAV) stars are short-period and low-amplitude pulsators with 10\,500--13\,000\,K effective temperatures, while their hotter siblings, the V777\,Her (or DBV) stars can be found at effective temperatures of roughly 22\,000--31\,000\,K. Both show light variations caused by non-radial \textit{g}-mode pulsations with periods between $\sim$100--1500\,s, typically with $\sim$mmag amplitudes. For reviews of the theoretical and observational aspects of pulsating white dwarf studies, see the papers of \citet{2008PASP..120.1043F}, \citet{2008ARA&A..46..157W}, \citet{2010A&ARv..18..471A}, and the recent review on ZZ\,Ceti pulsations based on \textit{Kepler} observations of \citet{2017ApJS..232...23H}. We also refer to the theoretical work of \citet{2013ApJ...762...57V}, concerning the reconstruction of the boundaries of the empirical ZZ\,Ceti instability strip up to the domain of the extremely low-mass DA pulsators. In addition, we also mention the observations of the so-called `outburst' events, which means recurring increases in the stellar flux in DAV stars being close to the red edge of the instability strip (see e.g. \citealt{2017ASPC..509..303B}).


\section{Observations and data analysis}

We performed the observations with the 1-m Ritchey-Chr\'etien-Coud\'e telescope located at Piszk\'estet\H o mountain station of Konkoly Observatory, Hungary. We obtained data with an Andor iXon+888 EMCDD and an FLI Proline 16803 CCD camera in white light. The exposure times were between 5 and 40\,s, depending on the weather conditions and the brightness of the target. 

Raw data frames were treated the standard way utilizing \textsc{iraf}\footnote{\textsc{iraf} is distributed by the National Optical Astronomy Observatories, which are operated by the Association of Universities for Research in Astronomy, Inc., under cooperative agreement with the National Science Foundation.} tasks: they were bias, dark and flat corrected before performing aperture photometry of field stars and low-order polynomial-fitting to the resulting light curves correcting for low-frequency atmospheric and instrumental effects. This latter smoothing of the light curves did not affect the known frequency domain of pulsating DA and DB stars. The comparison stars for the differential photometry were checked for variability or any kind of instrumental effects. Then, we converted the observational times of every data point to barycentric Julian dates in barycentric dynamical time ($\mathrm{BJD_{TDB}}$) using the applet of \citet{2010PASP..122..935E}\footnote{http://astroutils.astronomy.ohio-state.edu/time/utc2bjd.html}.    

We analysed the daily measurements with the command-line light curve fitting program \textsc{LCfit} \citep{2012KOTN...15....1S}. \textsc{LCfit} has linear (amplitudes and phases) and nonlinear (amplitudes, phases and frequencies) least-squares fitting options, utilizing an implementation of the Levenberg-Marquardt least-squares fitting algorithm. The program can handle unequally spaced and gapped datasets and is scriptable easily.

We performed the standard Fourier analyses of the whole dataset on WD~1310+583 with the photometry modules of the Frequency Analysis and Mode Identification for Asteroseismology (\textsc{famias}) software package \citep{2008CoAst.155...17Z}. Remaining traditional (see \citealt{1993A&A...271..482B}), we accepted a frequency peak as significant if its amplitude reached the 4 signal-to-noise ratio (S/N).

\subsection{Target selection strategy}

Our list of targets selected for observations is primarily based on the list of DAV and DBV candidate variables collected by the \textit{TESS} Compact Pulsators Working Group No.~8 (WG\#8) proposed for 20\,s cadence observations. This list of targets were compiled by the Monteral White Dwarf Database (MWDD)\footnote{http://dev.montrealwhitedwarfdatabase.org/home.html}, which is a collection of white dwarf stars with their physical parameters, in many cases, originate from different authors. Thus, it presents an overall view on a target queried, including their atmospheric parameters, coordinates and brightness values in different bandpasses, and an optical spectrum. For a more detailed description of the MWDD, we refer to the paper of \citet{2017ASPC..509....3D}.

We chose variable candidates close to the DAV or DBV empirical instability strips considering their effective temperature ($T_{\rmn{eff}}$) and surface gravity ($\mathrm{log}\,g$), and altogether, 14 white dwarf variable candidates were observed in the March--November, 2017 term. 

\section{Stars showing no light variations}

\begin{table*}
\centering
\caption{Summary of our observations of NOV stars performed at Piszk\'estet\H oi mountain station. `Exp' is the integration time used, \textit{N} is the number of data points and $\delta T$ is the length of the dataset including gaps. In the comment section, we list the 4$\langle {\rm A}\rangle$ significance levels in parentheses in mmag unit.}
\label{tabl:lognov}
\small
\begin{tabular}{lrccrrl}
\hline
Run & UT Date & Start time & Exp. & \textit{N} & $\delta T$ & Comment\\
 & (2017) & (BJD-2\,450\,000) & (s) &  & (h) & \\
\hline
\multicolumn{6}{l}{EGGR\,116:} & NOV(1)\\
01 & Mar 21 & 7834.377 & 30 & 1505 & 6.70 &\\
02 & Mar 30 & 7843.371 & 10 & 2350 & 6.54 & \\
\\
\multicolumn{6}{l}{EGGR\,162:} & NOV(1)\\
01 & Sep 15 & 8012.298 & 10 & 807 & 2.97 &\\
\\
\multicolumn{6}{l}{EGGR\,311:} & NOV(2)\\
01 & Nov 14 & 8072.259 & 30 & 679 & 6.28 & \\
02 & Nov 22 & 8080.188 & 30 & 384 & 3.55 & \\
\\
\multicolumn{6}{l}{GD\,190:} & NOV(4)\\
01 & Apr 25 & 7869.336 & 30 & 409 & 4.07 & \\
\\
\multicolumn{6}{l}{GD\,426:} & NOV(2)\\
01 & Oct 20 & 8047.431 & 30 & 591 & 5.44 &\\
\\
\multicolumn{6}{l}{GD\,83:} & NOV(2)\\
01 & Nov 22 & 8080.339 & 30 & 964 & 8.81 &\\
\\
\multicolumn{6}{l}{HG\,8-7:} & NOV(2)\\
01 & Oct 21 & 8048.480 & 15 & 588 & 3.32 & \\
\\
\multicolumn{6}{l}{PG\,1026+024:} & NOV(3)\\
01 & Mar. 16 & 7829.324 & 10 & 1078 & 4.26 &\\
\\
\multicolumn{6}{l}{WD\,0129+458:} & NOV(1)\\
01 & Oct 19 & 8046.221 & 30 & 1131 & 10.84 &\\ 
02 & Oct 31 & 8058.353 & 15 & 502 & 5.52 &\\
03 & Nov 15 & 8073.184 & 30 & 1209 & 11.07 &\\
\\
\multicolumn{6}{l}{WD\,0145+234:} & NOV(2)\\
01 & Oct 20 & 8047.219 & 30 & 533 & 4.94 & \\
\\
\multicolumn{6}{l}{WD\,0449+252:} & NOV(2)\\
01 & Oct 29 & 8056.401 & 40 & 294 & 3.72 &\\ 
02 & Oct 30 & 8057.321 & 30 & 827 & 8.79 &\\
\\
\multicolumn{6}{l}{WD\,0454+620:} & NOV(1)\\
01 & Nov 15 & 8072.529 & 10 & 1108 & 4.02 &\\
\hline
\end{tabular}
\end{table*}

\begin{table}
\centering
\caption{Summary of our observations performed at Piszk\'estet\H oi mountain station on the new light variable WD stars. `Exp' is the integration time used, \textit{N} is the number of data points and $\delta T$ is the length of the data sets including gaps.}
\label{tabl:logwd}
\small
\begin{tabular}{lrccrr}
\hline
Run & UT Date & Start time & Exp. & \textit{N} & $\delta T$ \\
 & (2017) & (BJD-2\,450\,000) & (s) &  & (h) \\
\hline
\multicolumn{6}{l}{WD 1310+583:} \\
01 & Mar 31 & 7844.296 & 10 & 2778 & 8.40 \\
02 & Apr 24 & 7868.332 & 10 & 782 & 3.13 \\
03 & Jul 13 & 7948.331 & 20 & 831 & 5.48 \\
04 & Jul 14 & 7949.405 & 30 & 388 & 3.61 \\
05 & Jul 16 & 7951.316 & 30 & 618 & 5.67 \\
06 & Jul 17 & 7952.342 & 20 & 465 & 2.98 \\
07 & Jul 18 & 7953.326 & 20 & 824 & 5.23 \\
08 & Jul 19 & 7954.314 & 20 & 705 & 4.53 \\
\multicolumn{3}{l}{Total:} & & \multicolumn{1}{r}{7\,391} & 39.02\\
\\
\multicolumn{6}{l}{EGGR\,120:} \\
01 & Apr 3 & 7847.407 & 10 & 2002 & 5.56 \\
\hline
\end{tabular}
\end{table}

\begin{table*}
\centering
\caption{WD\,1310+583: result of the Fourier-analysis of the whole dataset. The frequencies are listed in the order of the pre-whitening procedure. Error value for $f_2$ is the standard deviation of the Gaussian fitted to the peaks found around this frequency, while in the other cases errors were derived by Monte Carlo simulations.}
\label{tabl:wdfreq}
\begin{tabular}{lrrrrrrr}
\hline
 & \multicolumn{1}{c}{\textit{f}} & \multicolumn{1}{c}{\textit{$\delta$f}} & \multicolumn{1}{c}{\textit{P}} & Ampl. & Phase & S/N & Comment\\
 & \multicolumn{2}{c}{[$\mu$Hz]} & \multicolumn{1}{c}{[s]} & [mmag] & [$2\pi$] & & \\
\hline
$f_1$	&1439.858 & 0.001	&694.51	&12.8	&0.63 & 16.4 & \\
$f_2$	&1063.102 & 45	&940.64	& --	& -- & -- & \\
$f_3$	&1451.478 & 0.003	&688.95	&8.1	&0.87 & 10.5 & $\sim f_1$+1\,d$^{-1}$\\
$f_4$	&1958.452 & 0.001	&510.61	&6.9	&0.16 & 9.8 & \\
$f_5$	&1751.226 & 0.009	&571.03	&6.3	&0.79 & 8.8 & \\
$f_6$   &2739.544 & 0.002	&365.02	&6.3	&0.79 & 10.2 & \\
$f_7$	&963.032 & 0.003	&1038.39	&7.0	&0.86 & 9.9 & \\
$f_8$	&967.347 & 0.004	&1033.76	&5.9	&0.27 & 8.1 & close to $f_7$\\
$f_9$		&1751.155 & 0.010	&571.05	&5.4	&0.88 & 7.6 & close to $f_5$\\
$f_{10}$	&1469.611 & 0.003	&680.45	&3.6	&0.69 & 4.6 & close to $f_1$\\
$f_{11}$	&1037.324 & 0.002	&964.02	&5.0	&0.72 & 5.7 & close to $f_2$\\
$f_{12}$	&633.984 & 0.005	&1577.33	&6.6	&0.11 & 8.1 & \\
$f_{13}$	&632.088 & 0.006	&1582.06	&5.6	&0.83 & 6.8 & close to $f_{12}$\\
$f_{14}$	&2767.140 & 0.003	&361.38	&3.0	&0.91 & 4.8 & close to $f_6$\\
$f_{15}$	&2890.388 & 0.003	&345.97	&2.6	&0.17 & 4.7 & $2f_1$+1\,d$^{-1}$\\
$f_{16}$	&3197.877 & 0.004	&312.71	&2.3	&0.22 & 4.3 & $\sim f_1+f_5$\\
$f_{17}$	&4493.635 & 0.005	&222.54	&1.6	&0.97 & 4.4 & $\sim f_5 + f_6$\\
\hline
\end{tabular}
\end{table*}

\begin{table*}
\centering
\caption{Best-fitting models for WD\,1310+583 derived by seven periods (first two rows) and nine periods (third row).}
\label{tabl:models}
\small
\begin{tabular}{lrrp{8mm}p{8mm}p{8mm}p{8mm}p{9mm}p{9mm}p{8mm}p{8mm}l}
\hline
$T_{\rmn{eff}}$ (K) & \multicolumn{1}{c}{$M_*/M_{\sun}$} & \multicolumn{1}{c}{-log\,$M_\rmn{H}$} & 
\multicolumn{8}{c}{Periods in seconds ($l$)} & $\sigma_\mathrm{{rms}}$\\
\hline 
10\,400 & 0.78 & 8.2 & 364.9(1) & 510.6(2) & 569.9(1) & 694.2(2) & 1038.7(2) & 1579.8(1) & & & 1.1\\
11\,600 & 0.74 & 4.0 & 364.5(1) & 508.9(2) & 568.9(1) & 694.5(1) & 1037.1(2) & 1576.2(1) & & & 1.3\\
11\,900 & 0.80 & 7.6 & 362.1(1) & 511.3(1) & 572.2(2) & 692.3(1) & 1038.2(1) & 1575.7(1) & 389.9(2)  & 545.7(2)  & 1.6\\  
\multicolumn{3}{l}{\textit{Observations:}} & 365.0 & 510.6 & 571.0 & 694.5 & 1038.4 & 1577.3 & 390.9 & 545.5 & \\ 
\hline
\end{tabular}
\end{table*}

\begin{table}
\centering
\caption{Physical parameters of the 14 targets observed in our survey. We denoted by \textit{G}
at the surface gravity when the source of the original physical parameters are from the database of \citet{2011ApJ...743..138G}. We corrected these $T_{\mathrm{eff}}$ and $\mathrm{log}\,g$ values according to the findings of \citet{2013A&A...559A.104T} based on radiation-hydrodynamics three-dimensional simulations of convective DA stellar atmospheres. In the other cases the source of the parameters was either \citet{2015ApJS..219...19L} ($L$) or \citet{2011ApJ...737...28B} ($B$), respectively.}
\label{tabl:tefflogg}
\small
\begin{tabular}{lrccl}
\hline
ID & Spectral type & $T_{\mathrm{eff}}$ & $\mathrm{log}\,g$ & \textit{TESS} mag. \\
 & & (K) & (dex) & \\
\hline
EGGR\,116 & DA & 13100 & 7.87$^G$ & 13.6($I$)\\
EGGR\,162 & DA & 13100 & 8.02$^G$ & 13.2\\
EGGR\,311 & DA & 12800 & 7.98$^G$ & 14.5\\
GD\,190 & DB & 22630 & 8.04$^B$ & 14.9($I$)\\
GD\,426 & DA & 10920 & 8.10$^G$ & 15.6\\
GD\,83 & DA & 10390 & 7.93$^G$ & 14.9($I$)\\
HG\,8-7 & DA & 12690 & 8.05$^G$ & 13.6($I$)\\
PG\,1026+024 & DA & 13110 & 8.06$^G$ & 14.1($V$)\\
WD\,0129+458 & DA & 10680 & 7.97$^L$ & 14.3($I$)\\
WD\,0145+234 & DA & 13000 & 8.13$^G$ & 14.2\\
WD\,0449+252 & DAH & 11500 & 8.00$^L$ & 14.9($V$)\\
WD\,0454+620 & DA+dM & 10960 & 8.88$^L$ & 12.3\\
\\
EGGR\,120 & DA & 10170 & 8.03$^G$ & 14.8\\
WD\,1310+583 & DA or & 10460 & 8.17$^G$ & 13.8\\
 & DA+DA & & & \\
\hline
\end{tabular}
\end{table}

\begin{figure*}
\centering
\includegraphics[width=0.65\textwidth, angle=270]{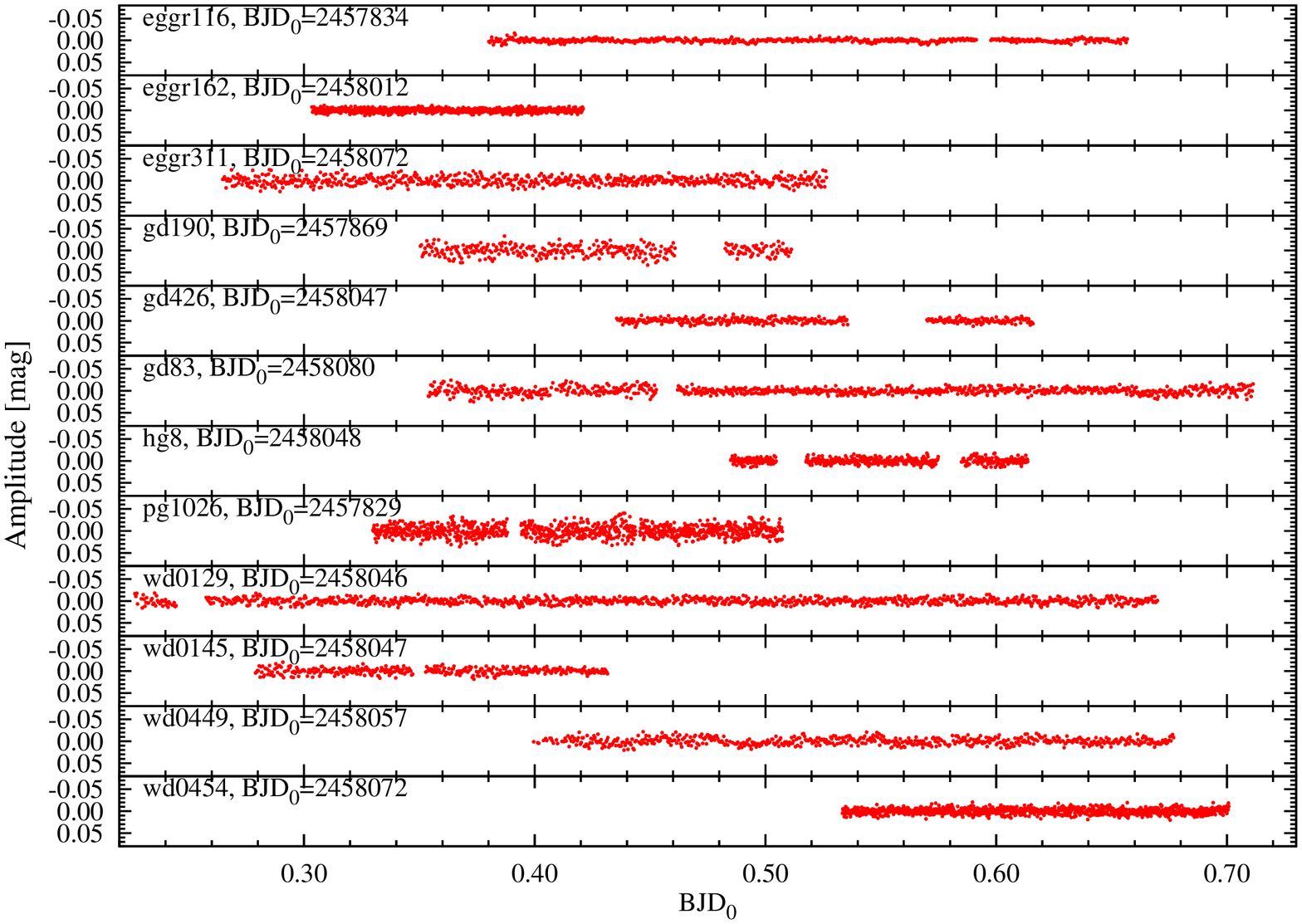}
\caption{Representative light curves of the NOV stars.}
\label{fig:lcnov}
\end{figure*}

12 out of the 14 stars in our sample were not observed to vary (NOV stars). We list the log of their observations in Table~\ref{tabl:lognov}. We did not find any significant frequencies in their Fourier transform which would suggest that pulsation operates in them. 

The significance levels for the different light curves were calculated by computing moving averages of the FTs of the measurements, which provided us an average amplitude level ($\langle {\rm A}\rangle$). If a target was observed on more than one night, we utilized the FT of all the available data. We considered a peak significant if it reached or exceeded the 4$\langle {\rm A}\rangle$ level. Table~\ref{tabl:lognov} also summarizes these 4$\langle {\rm A}\rangle$ significance levels in parentheses and in mmag units, found to be around 1--2\,mmag in most cases. 


We present representative light curves of these NOV stars and the FTs of the observations in Fig.~\ref{fig:lcnov} and in Appendix~\ref{app:novft}, respectively.  

\section{New variables}

We successfully identified two new DA white dwarf variables (EGGR~120 and WD~1310+583) in our sample, and three new light variables of other types amongst the field stars: a candidate delta Scuti/beta Cephei, a candidate W UMa-type star and an eclipsing binary.

Table~\ref{tabl:logwd} shows the journal of observations of the new WD light variables.

\subsection{EGGR~120}

\begin{figure*}
\centering
\includegraphics[width=0.9\textwidth]{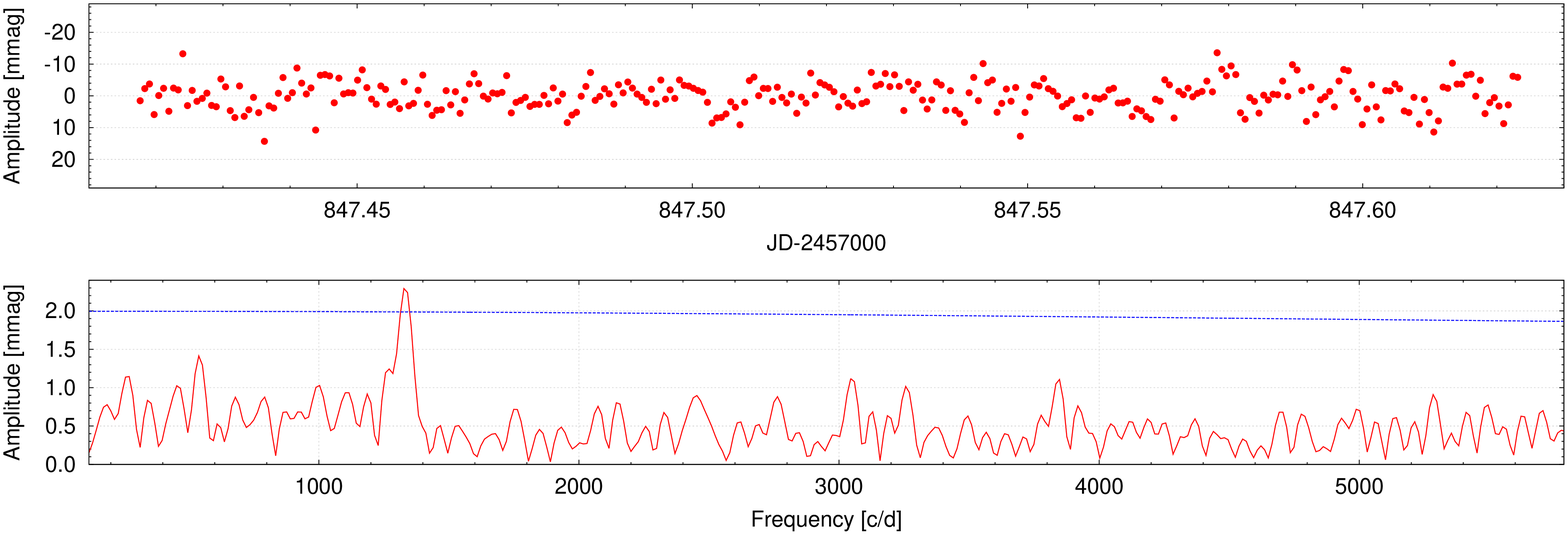}
\caption{EGGR~120: light curve its Fourier transform. Blue line denotes the 4$\langle {\rm A}\rangle$ significance level.\label{fig:eggr120}}
\end{figure*}

EGGR~120 ($V=14.8$\,mag, $\alpha_{2000}=16^{\mathrm h}39^{\mathrm m}28^{\mathrm s}$, $\delta_{2000}=+33^{\mathrm d}25^{\mathrm m}22^{\mathrm s}$) was found to be a light variable by one night of observation.

Figure~\ref{fig:eggr120} shows its light curve and the corresponding FT. We detected only one significant frequency at 1332\,$\mu$Hz with 2.3\,mmag amplitude. This frequency value corresponds to $\sim751$\,s periodicity, which places this star in the class of cool DAV stars \citep{2006ApJ...640..956M}, in agreement with its low estimated effective temperature. We plan to observe the star during the next observing season to obtain a more complete picture on its pulsation properties. 

\subsection{WD~1310+583}

\begin{figure*}
\centering
\includegraphics[width=0.65\textwidth, angle=270]{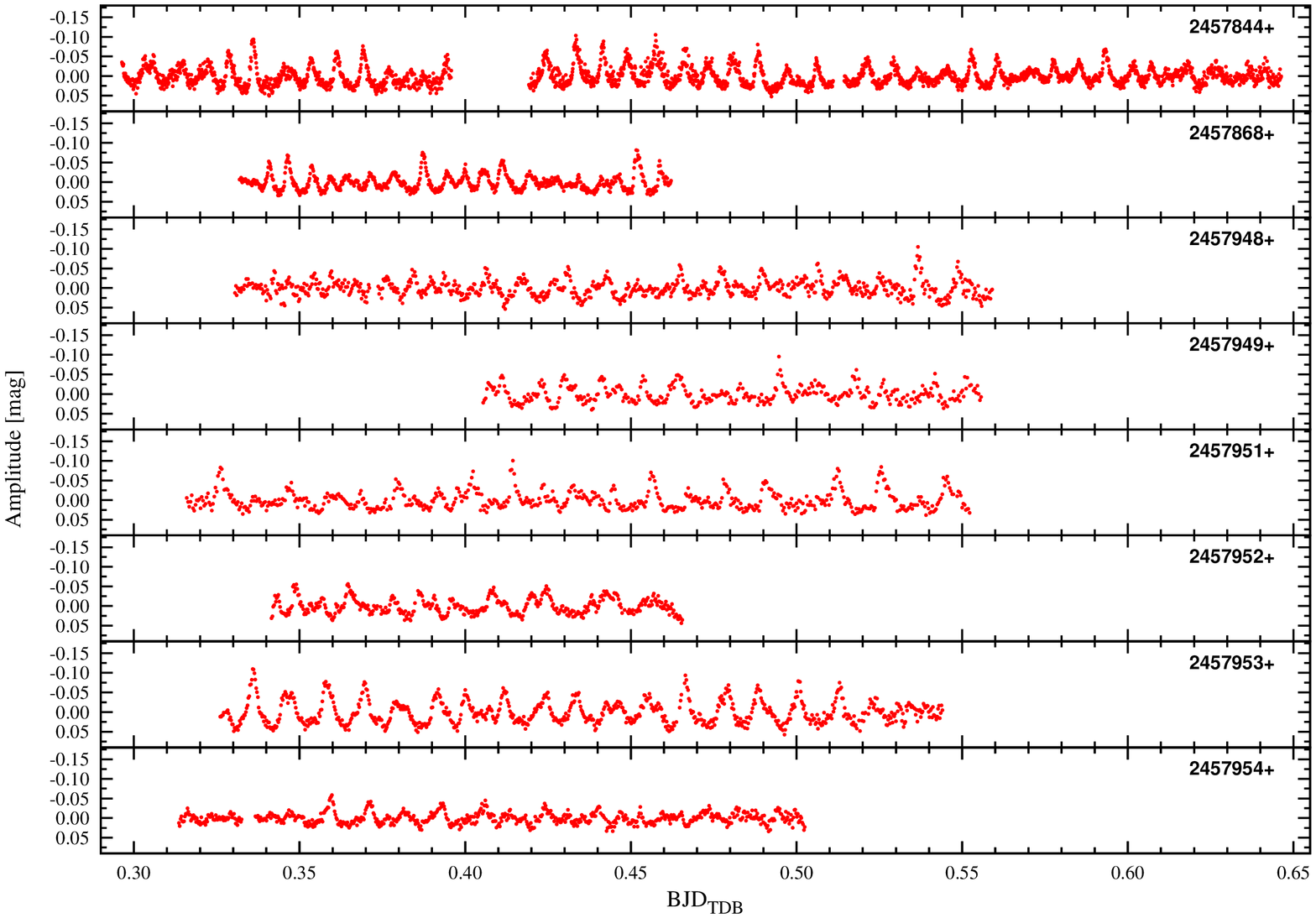}
\caption{Normalized differential light curves of the observations of WD\,1310+583.}{\label{fig:wd1310lc}}
\end{figure*}

\begin{figure*}
\centering
\includegraphics[width=0.9\textwidth]{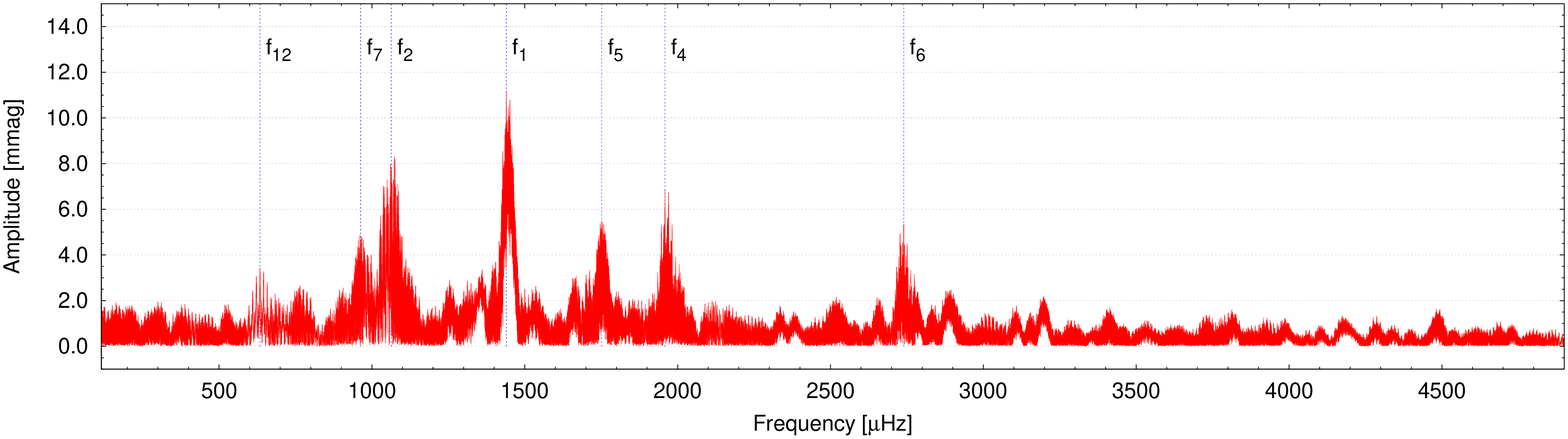}
\caption{Fourier transform of the whole dataset obtained on WD\,1310+583. We marked the frequencies that can be regarded as independent pulsation modes with blue dashed lines (cf. Table~\ref{tabl:wdfreq}).}{\label{fig:ft}}
\end{figure*}

WD~1310+583 ($B=13.9$\,mag, $\alpha_{2000}=13^{\mathrm h}12^{\mathrm m}58^{\mathrm s}$, $\delta_{2000}=+58^{\mathrm d}05^{\mathrm m}11^{\mathrm s}$) was observed on eight nights during the March--July, 2017 term. We performed most of the measurements in July, when during a one-week observing session six out of the seven nights were clear.

Figure~\ref{fig:wd1310lc} and Appendix~\ref{app:wdft} shows the light curves of observations and their FTs, respectively.

Considering the FTs, it is conspicuous that the amplitudes of frequencies vary from night-to-night. It might indicate real amplitude variations, that is, the energy content of some frequencies may vary in short time scales. Amplitude and phase variations are well-known phenomena amongst pulsating white dwarf stars, observed both from the ground (see e.g. the short overview of \citealt{2003ASPC..292..247H}) and from space (e.g. GD\,1212, \citealt{2014ApJ...789...85H}). However, we have to be cautious in the interpretation of these phenomena as true changes in the amplitudes and phases, as the beating of closely spaced frequencies could also be a possible explanation.

The Fourier analysis of the whole dataset resulted in the determination of 17 frequencies, which are listed in Table~\ref{tabl:wdfreq}.

Considering the frequencies in Table~\ref{tabl:wdfreq}, we determined closely spaced peaks with frequency separations of 0.08, 4.6 and 4.7\,$\mu$Hz. There are at least two possibilities: these are the results of short-term amplitude variations, or in the case of the similar, 4.6 and 4.7\,$\mu$Hz separations, rotationally split frequencies can be found at these domains. 

There are also a bit widely spaced doublets, which could also originate from rotational splitting of frequencies. Their separations are 25.8, 27.6 and 29.8\,$\mu$Hz, that is, close to each other. Assuming that these are rotationally split $l=1$ frequencies ($m=0,1$ or $m=-1,0$ pairs), the star's rotational period may be around 5\,h, applying the equation as follows: 

\begin{equation}
\label{eq:rot}
\delta f_{k,\ell,m} = \delta m (1-C_{k,\ell}) \Omega, 
\end{equation}
\noindent where the coefficient \mbox{$C_{k,\ell} \approx 1/\ell(\ell+1)$} for high-overtone ($k\gg\ell$) $g$-modes and 
$\Omega$ is the (uniform) rotation frequency.

Considering the 4.6 and 4.7\,$\mu$Hz separations, the star's rotational period could be 1.3\,d. As both rotation period values are acceptable for a white dwarf (see e.g. \citealt{2008PASP..120.1043F} or \citealt{2017ApJS..232...23H}), we cannot decide yet, which frequency separations we should consider as results of rotational splitting, if any. We also note that the daily (1\,d$^{-1}=11.6\,\mu$Hz) alias problem also makes the determination of independent pulsation modes difficult.

In the case of the peaks around 1060\,$\mu$Hz, we found many closely spaced frequencies in the FT of the whole dataset. Thus, we fitted the peaks with a Gaussian and decided to list the resulting maximum frequency and the function's standard deviation at $f_2$ in Table~\ref{tabl:wdfreq}. 

Finally, the frequencies we regard as independent modes are the following seven frequencies out of the 17 determined: $f_1$, $f_2$, $f_4$, $f_5$, $f_6$, $f_7$ and $f_{12}$.
We denoted them in the Fourier transform of the whole dataset in Fig.~\ref{fig:ft}. Considering the large uncertainty in the frequency determination of $f_2$, we did not use it as an input for asteroseismic investigations, but performed asteroseismic fits with the six remaining modes.


\subsubsection{Preliminary asteroseimology}

We built a model grid for the preliminary asteroseismic investigations of WD\,1310+583 utilizing the White Dwarf Evolution Code (\textsc{wdec}; \citealt{1969ApJ...156.1021K, 1974PhDT........56L, 1975ApJ...200..306L, 1991PhDT.........XX, 1986PhDT.........2K, 1990PhDT.........5W, 1993PhDT.........4B, 1998PhDT........21M, 2008ApJ...675.1512B}).

The \textsc{wdec} evolves a hot polytrope model ($\sim10^5$\,K) down to the requested temperature, and provides an equilibrium, thermally relaxed solution to the stellar structure equations. Then we are able to calculate the set of possible eigenmodes according to the adiabatic equations of non-radial stellar oscillations \citep{1989nos..book.....U}.

We utilized the integrated evolution/pulsation form of the \textsc{wdec} code created by \citet{2001PhDT.........1M} to derive the pulsation periods for the models with the given stellar parameters. 

We calculated the periods of dipole ($l=1$) and quadrupole ($l=2$) modes for the model stars considering the limited visibility of high spherical degree ($l$) modes due to geometric cancellation effects. The goodness of the fit between the observed ($P_i^{\rmn{obs}}$) and calculated ($P_i^{\rmn{calc}}$) periods was characterized by the root mean square ($\sigma_\mathrm{{rms}}$) value calculated for every model with the \textsc{fitper} program of \citet{2007PhDT........13K}:

\begin{equation}
\sigma_\mathrm{{rms}} = \sqrt{\frac{\sum_{i=1}^{N} (P_i^{\rmn{calc}} - P_i^{\rmn{obs}})^2}{N}}
\label{equ1}
\end{equation}

\noindent where \textit{N} is the number of observed periods.

We built our model grid using the core composition profiles of \citet{1997ApJ...486..413S} based on evolutionary calculations. We varied three input parameters of the WDEC: $T_{\rmn{eff}}$, $M_*$ and $M_\rmn{H}$. The grid covers the parameter range $10\,200-12\,000$\,K in $T_{\rmn{eff}}$, $0.55-0.85\,M_{\sun}$ in stellar mass, $10^{-4}-10^{-9}\,M_*$ in $M_\rmn{H}$, and we fixed the mass of the helium layer at the theoretical maximum value of $10^{-2}\,M_*$. 
We used step sizes of $100$\,K ($T_{\rmn{eff}}$), $0.01\,M_{\sun}$ ($M_*$) and $0.2$\,dex (log\,$M_\rmn{H}$). 

The mass of WD\,1310+583 ($\mathrm{log}\,g=8.17$ by optical spectroscopy) was determined utilizing the theoretical masses calculated for DA stars by \citet{1996ApJ...468..350B}, which resulted $0.7\,M_{\sun}$ for the star. The error of the mass determination is about $0.03\,M_{\sun}$. The effective temperature of the star is derived to be 10\,460\,K with about $200$\,K uncertainty value in the optical. However, based on far-ultaviolet (FUV) spectroscopy, it turned out that WD\,1310+583 may be a double degenerate binary system, in which one component with about 11\,600\,K effective temperature is close to the middle of the ZZ Ceti instability strip, while the other component may be much cooler, about $T_{\rmn{eff}}=7900$\,K \citep{2018MNRAS.473.3693G}. Furthermore, utilizing the time-tag information in the FUV spectrum, \citet{2018MNRAS.473.3693G} determined two pulsation frequencies of WD\,1310+583 at 391 and 546\,s, respectively. We did not find them in our measurements, thus these may represent new pulsation frequencies besides our findings.


We found, that utilizing our model grid, the best-fitting model (model with the lowest $\sigma_\mathrm{{rms}}$ value) has stellar mass higher than the value determined by optical spectroscopy ($0.78\,M_{\sun}$), but its effective temperature is close to the value calculated from the optical spectrum ($10\,400$\,K). However, note that in this case the dominant mode is $l=2$. Assuming that at least four of the modes is $l=1$ (including the dominant frequency), considering the better visibility of $l=1$ modes over $l=2$ ones, the best-fitting model has $T_{\rmn{eff}}=11\,600$\,K and $M_*=0.74\,M_{\sun}$ ($\sigma_\mathrm{{rms}}$=1.3\,s). That is, this solution has stellar mass close to the value determined by optical spectroscopy, but its effective temperature fits better to the value calculated from the FUV fitting.

We tried another fit adding the two frequencies found by \citet{2018MNRAS.473.3693G}, that is, we fitted eight periods with the calculated ones. In this case, the best-fitting model has $T_{\rmn{eff}}=11\,900$\,K and $M_*=0.80\,M_{\sun}$ ($\sigma_\mathrm{{rms}}$=1.6\,s). This is also the best-fitting model assuming that at least five of the modes is $l=1$.

We summarized our model findings in Table~\ref{tabl:models}. Considering the effective temperatures of the best-fitting models, they seem to confirm the higher value determined by FUV observations. The relatively large amplitude pulsations also support the idea that the pulsating component of the WD\,1310+583 system may be closer to the middle of the ZZ Ceti instability strip, than it is at the red edge (cf. \citealt{2017ApJS..232...23H}).

\subsection{New variables of other types}

We identified an eclipsing binary on the field of WD\,0129+458, a delta Scuti/beta Cephei candidate on the field of WD\,0454+620, and a W UMa variable candidate on the field of GD\,83. Their distances to the white dwarfs are $\sim 4.4, 3.2$ and $1.5$\,arcmin, that is, they do not contaminate the large ($\sim 21$\,arcsec) \textit{TESS} pixels.

\begin{figure}
\centering
\includegraphics[width=0.48\textwidth]{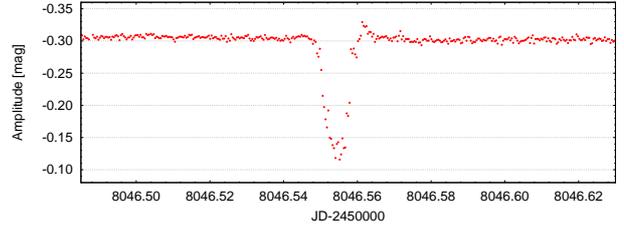}
\caption{Light curve of the eclipsing binary found on the field of WD\,0129+458.}{\label{fig:ecl}}
\end{figure}

\begin{figure}
\centering
\includegraphics[width=0.48\textwidth]{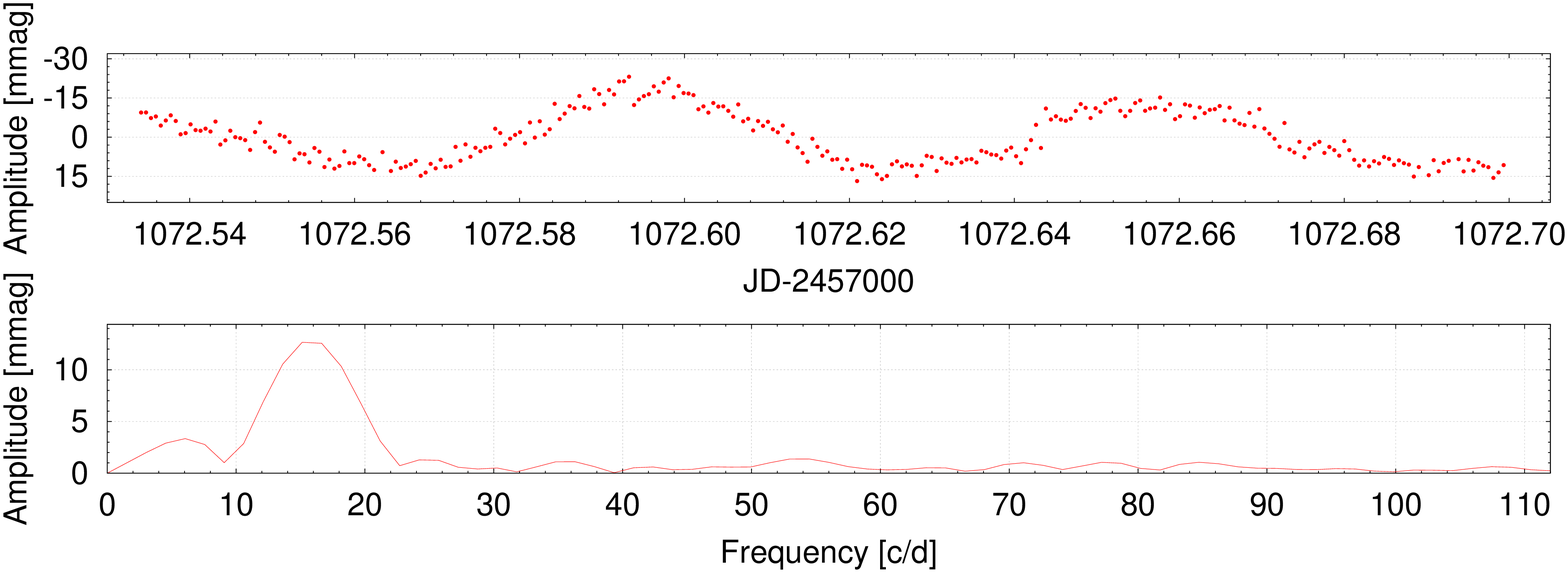}
\caption{Light curve and its Fourier transform of the variable found on the field of WD\,0454+620.}{\label{fig:dsct}}
\end{figure}

\begin{figure}
\centering
\includegraphics[width=0.48\textwidth]{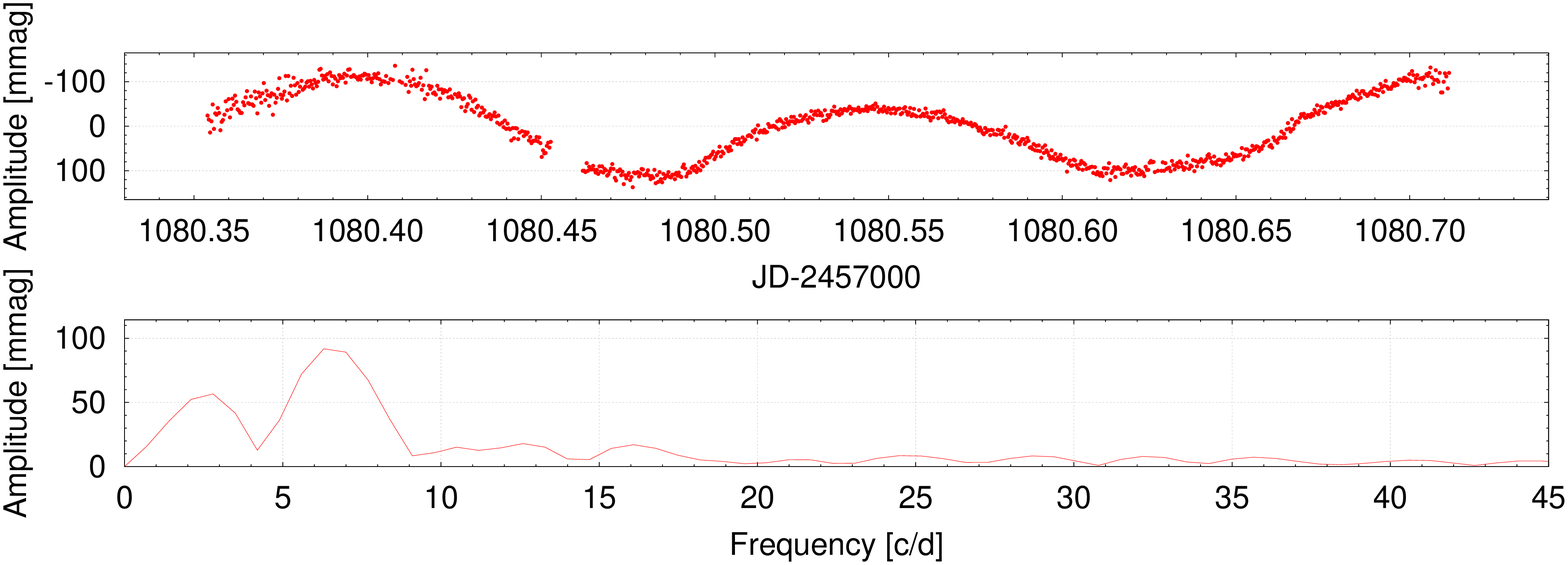}
\caption{Light curve and its Fourier transform of the WUMa-type variable candidate found on the field of GD\,83.}{\label{fig:wuma}}
\end{figure}

\subsubsection{Eclipsing binary}

We could observe one eclipse only (see Fig.~\ref{fig:ecl}). We tried to catch another minimum during further observations, but we did not succeed. Thus, we cannot tell at this moment, whether we saw a main or a secondary minimum in our light curve. The star's 2MASS identifier is 01323981+4600441 ($J=12.4$\,mag, $\alpha_{2000}=01^{\mathrm h}32^{\mathrm m}40^{\mathrm s}$, $\delta_{2000}=+46^{\mathrm d}00^{\mathrm m}44^{\mathrm s}$; \citealt{2003yCat.2246....0C}).

\subsubsection{Delta Scuti/beta Cephei variable candidate}

We found one significant frequency of the light variation of this star at 15.7\,d$^{-1}$ ($\sim$1.5\,h) with 13\,mmag amplitude (Fig.~\ref{fig:dsct}). This pulsational behaviour could be typical both for delta Scuti and beta Cephei stars, respectively. Colour or spectroscopical measurements could help to decide which type of pulsating variables this object belongs to, however, none of them are available at this moment. Its 2MASS identifier is 04584888+6212098 ($J=12.0$\,mag, $\alpha_{2000}=04^{\mathrm h}58^{\mathrm m}49^{\mathrm s}$, $\delta_{2000}=+62^{\mathrm d}12^{\mathrm m}10^{\mathrm s}$).  

\subsubsection{W UMa variable candidate}

Further finding is a W UMa-type variable candidate. Figure~\ref{fig:wuma} shows its light curve and Fourier spectrum. The dominant periodicity is at 6.75\,d$^{-1}$ ($\sim$0.15\,d). The star's 2MASS identifier is 07131730+2135152 ($J=14.5$\,mag, $\alpha_{2000}=07^{\mathrm h}13^{\mathrm m}17^{\mathrm s}$, $\delta_{2000}=+21^{\mathrm d}35^{\mathrm m}15^{\mathrm s}$).

\section{Summary and Conclusions}

\begin{figure*}
\centering
\includegraphics[width=0.75\textwidth]{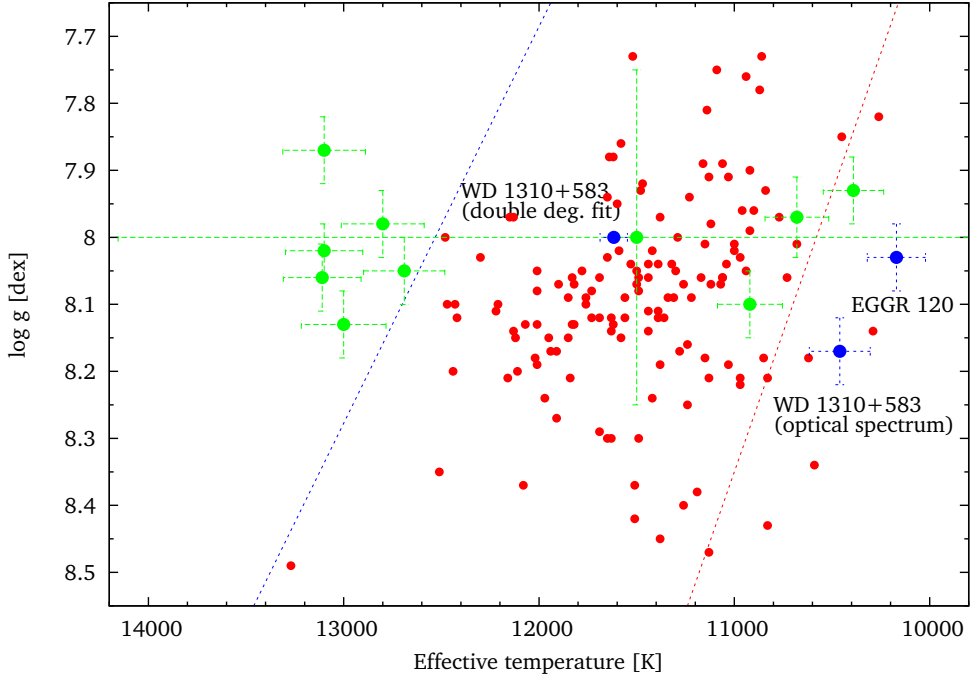}
\caption{Known variable stars (red filled dots) and the newly observed ZZ Ceti candidates and DAVs (green and blue dots, respectively) in the $T_{\mathrm{eff}}$ -- $\mathrm{log}\,g$ diagram. Blue and red dashed lines denote the hot and cool boundaries of the instability strip, according to \citet{2015ApJ...809..148T}. 
In the case of WD\,1310+583, the double degenerate solution assumes two white dwarfs with fixed $\mathrm{log}\,g=8.0$\,dex values. The effective temperature was determined by the simultaneous fitting of both the optical spectrum and the FUV to near-infrared photometric data with two white dwarf models \citep{2018MNRAS.473.3693G}.}
{\label{fig:instab}}
\end{figure*}

We aimed to perform survey observations to find new white dwarf pulsators for the \textit{TESS} mission. For this purpose, we collected photometric time-series data on 14 white dwarf variable-candidates at Konkoly Observatory during the March--November term in 2017. Besides the visual inspection of the light curves, we performed Fourier analysis of all datasets and successfully identified two new ZZ\,Ceti stars: EGGR\,120 and WD\,1310+583. In the case of EGGR\,120, which was observed on one night only, we found one significant frequency at 1332\,$\mu$Hz with 2.3\,mmag amplitude. We could observe WD\,1310+583 on eight nights altogether, and determined 17 significant frequencies by the whole dataset. Seven of them seems to be independent pulsation modes between 634 and 2740\,$\mu$Hz frequencies, and we performed preliminary asteroseismic investigations of the star utilizing six of these periods. 

We also identified three new variables on the fields of white dwarf candidates: an eclipsing binary, a delta Scuti/beta Cephei candidate and a W UMa-type star. The periods of their light variations are long enough that the 30\,min time sampling of the full-frame images (FFIs) of \textit{TESS} will be enough to study their pulsations and eclipses.

Figure~\ref{fig:instab} shows the classical ZZ Ceti instability strip with plots of the known DAV stars (red filled dots) and the stars presented in this paper (green and blue dots with errorbars, respectively). We collected the atmospheric parameters of known DAV stars utilizing the database of \citet{2016IBVS.6184....1B}, in which the authors listed corrected $T_{\mathrm{eff}}$ and $\mathrm{log}\,g$ values for the three-dimensional dependence of convection for most of the objects. In Table~\ref{tabl:tefflogg} we summarize the physical parameters of the 14 targets observed in our survey.

The newly discovered, relatively bright WD variables are excellent targets for small telescopes, especially WD\,1310+583, which shows larger amplitude light variations than EGGR\,120. Considering Fig.~\ref{fig:instab}, they lie close to the middle and the red edge of the ZZ Ceti instability domain, respectively, with good agreement with their pulsational properties: relatively long periods, non-sinusoidal light curves, and in the case of the longer observed WD\,1310+583, the closely spaced frequencies suggests the presence of amplitude and phase variations. With space photometry and additional ground-based follow-up observations planned, hopefully, we will learn much more about they pulsation behaviour in the near future.

Considering all the proposed white dwarf targets for \textit{TESS} observations, almost all of them have only been observed from the ground up to now. Sometimes these observations are even limited to the usually short discovery light curves. The 27-d or longer, uninterrupted \textit{TESS} measurements will outperform most of the available data on these bright pulsators. Data simulations also show, that in terms of signal-to-noise ratio, considering equal monitoring time, \textit{TESS} data are expected to be roughly equivalent to \textit{Kepler} data obtained for stars five magnitudes fainter. Besides, taking into account that pulsating white dwarfs down to magnitude 19 have successfully been observed with \textit{K2}, this suggests that \textit{TESS} can provide useful data at least down to magnitude $\sim 15$.

\section*{Acknowledgements}
The authors thank the anonymous referee for the constructive comments and recommendations on the manuscript.
The authors thank Agn\`es Bischoff-Kim for providing her version of the \textsc{wdec} and the \textsc{fitper} program.
\'AS was supported by the J\'anos Bolyai Research Scholarship of the Hungarian Academy of Sciences, and he also acknowledges the financial support of the Hungarian NKFIH Grant K-113117. \'AS and ZsB acknowledges the financial support of the Hungarian NKFIH Grants K-115709 and K-119517. ZsB acknowledges the support provided from the National Research, Development and Innovation Fund of Hungary, financed under the PD\_17 funding scheme, project no. PD-123910. This project has been supported by the Lend\"ulet grant LP2012-31 of the Hungarian Academy of Sciences and by the GINOP-2.3.2-15-2016-00003 grant of the Hungarian National Research, Development and Innovation Office (NKFIH).
Support for this work was provided by NASA through Hubble Fellowship grant \#HST-HF2-51357.001-A, awarded by the Space Telescope Science Institute, which is operated by the Association of Universities for Research in Astronomy, Incorporated, under NASA contract NAS5-26555.




\bibliographystyle{mnras}
\bibliography{tess} 




\appendix
\section{}

\begin{figure*}
\centering
\includegraphics[width=0.75\textwidth]{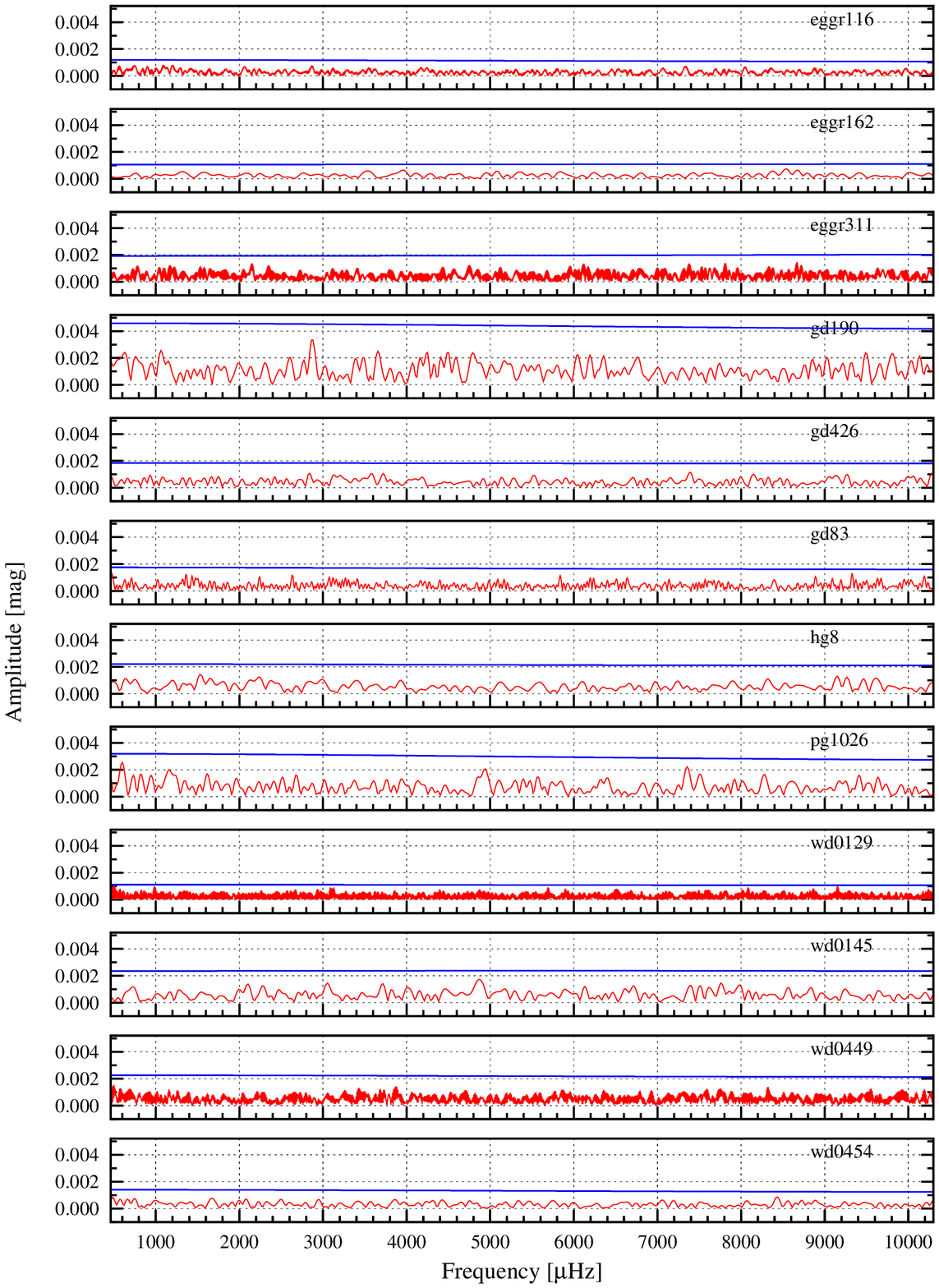}
\caption{Fourier transforms of the light curves of NOV stars. Blue lines denote the 4$\langle {\rm A}\rangle$ significance level for the detection of possible pulsation frequencies.}{\label{app:novft}}
\end{figure*}

Figure~\ref{app:novft}: Fourier transforms of the light curves of NOV stars. Blue lines denote the 4$\langle {\rm A}\rangle$ significance level for the detection of possible pulsation frequencies.

\section{}

\begin{figure*}
\centering
\includegraphics[width=0.75\textwidth]{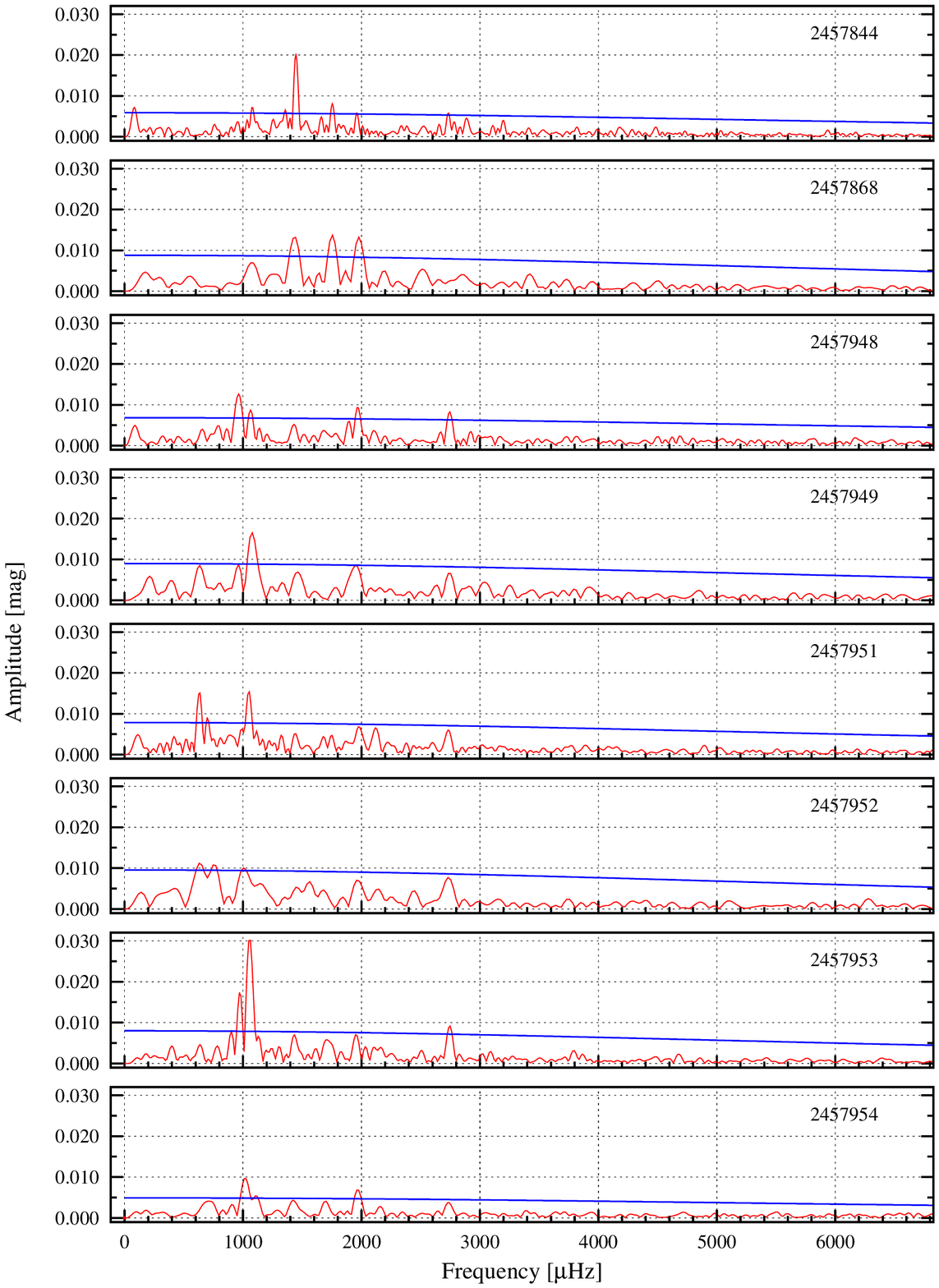}
\caption{Fourier transforms of the nightly observations of WD\,1310+583.}{\label{app:wdft}}
\end{figure*}

Figure~\ref{app:wdft}: Fourier transforms of the nightly observations of WD\,1310+583. Blue lines denote the 4$\langle {\rm A}\rangle$ significance levels.


\bsp	
\label{lastpage}
\end{document}